\documentclass[vecphys]{svmult}

\usepackage{color}
\usepackage{makeidx}         
\usepackage{graphicx}        
\usepackage{multicol}        
\usepackage{cite}            
\usepackage[bottom]{footmisc}

\makeindex             

\begin{document}

\title{Magnetic Avalanches in Molecular Magnets}
\author{Myriam P. Sarachik}
\institute{City College of New York, CUNY, New York, NY 10031, USA
\texttt{sarachik@sci.ccny.cuny.edu}}

\maketitle

\begin{abstract}
The reversal of the magnetization of crystals of molecular magnets that have a large spin and high anisotropy barrier generally proceeds below the blocking temperature by quantum tunneling.  This is manifested as a series of controlled steps in the hysteresis loops at resonant values of the magnetic field where energy levels on opposite sides of the barrier cross. An abrupt reversal of the magnetic moment of the entire crystal can occur instead by a process commonly referred to as a magnetic avalanche, where the molecular spins reverse along a deflagration front that travels through the sample at subsonic speed.  In this chapter, we review experimental results obtained to date for magnetic deflagration in molecular nanomagnets.
\end{abstract}

\noindent
\section{Background}\label{sec1}
First reported by Paulsen and Park\cite{Paulsen}, magnetic avalanches occur in many different molecular magnets.  Systematic experimental studies of avalanches have focussed largely on crystals of  Mn$_{12}$-ac [Mn$_{12}$O$_{12}$(CH$_3$COO)$_{16}$(H$_2$O)$_4$] a particularly simple, high-symmetry prototype of this class of materials.

\begin{figure}[h]
\vspace{-1in}
 \includegraphics[width=.37\textwidth]{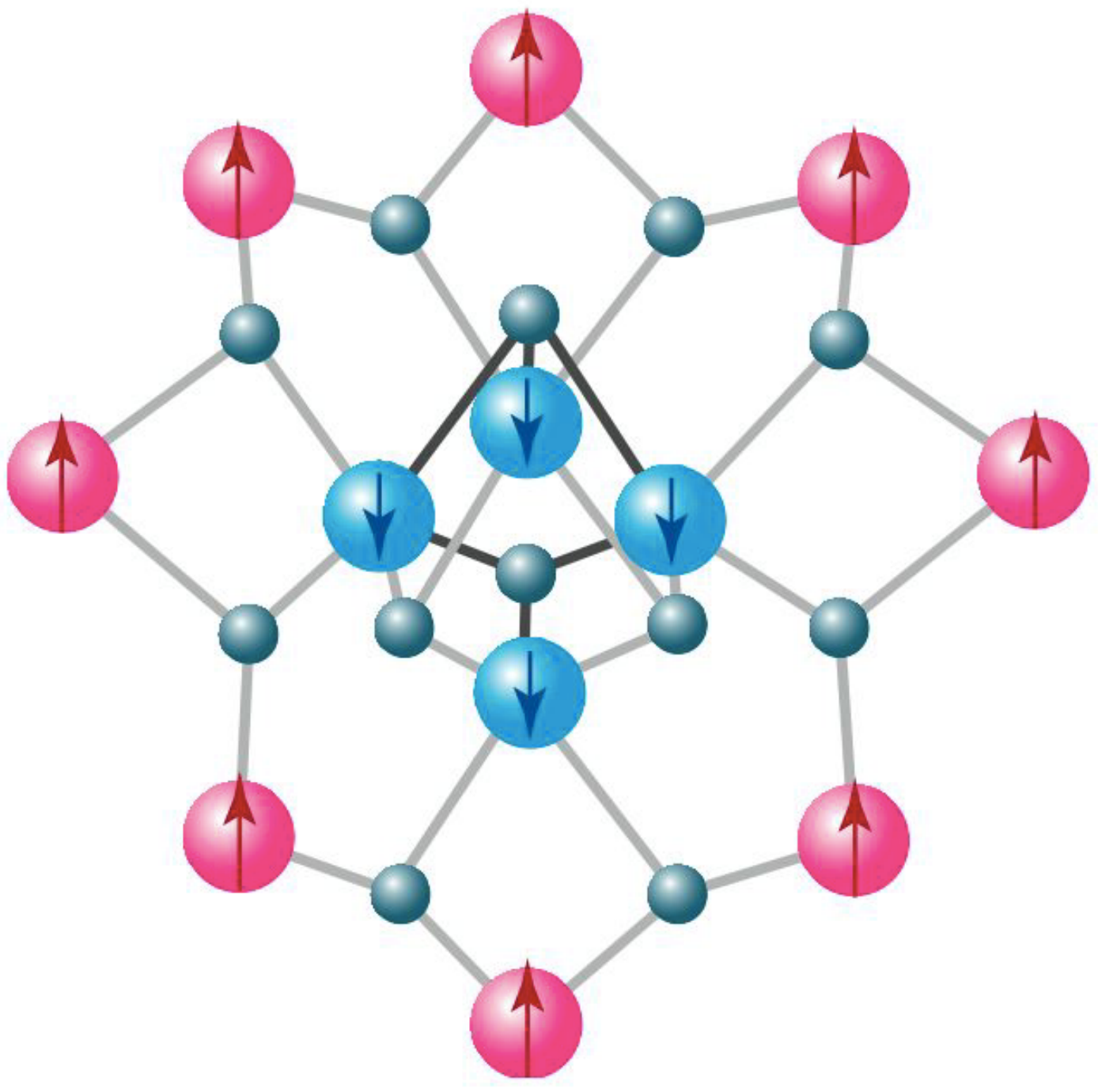}
\hspace{-.2in}
\includegraphics[width=.39\textwidth]{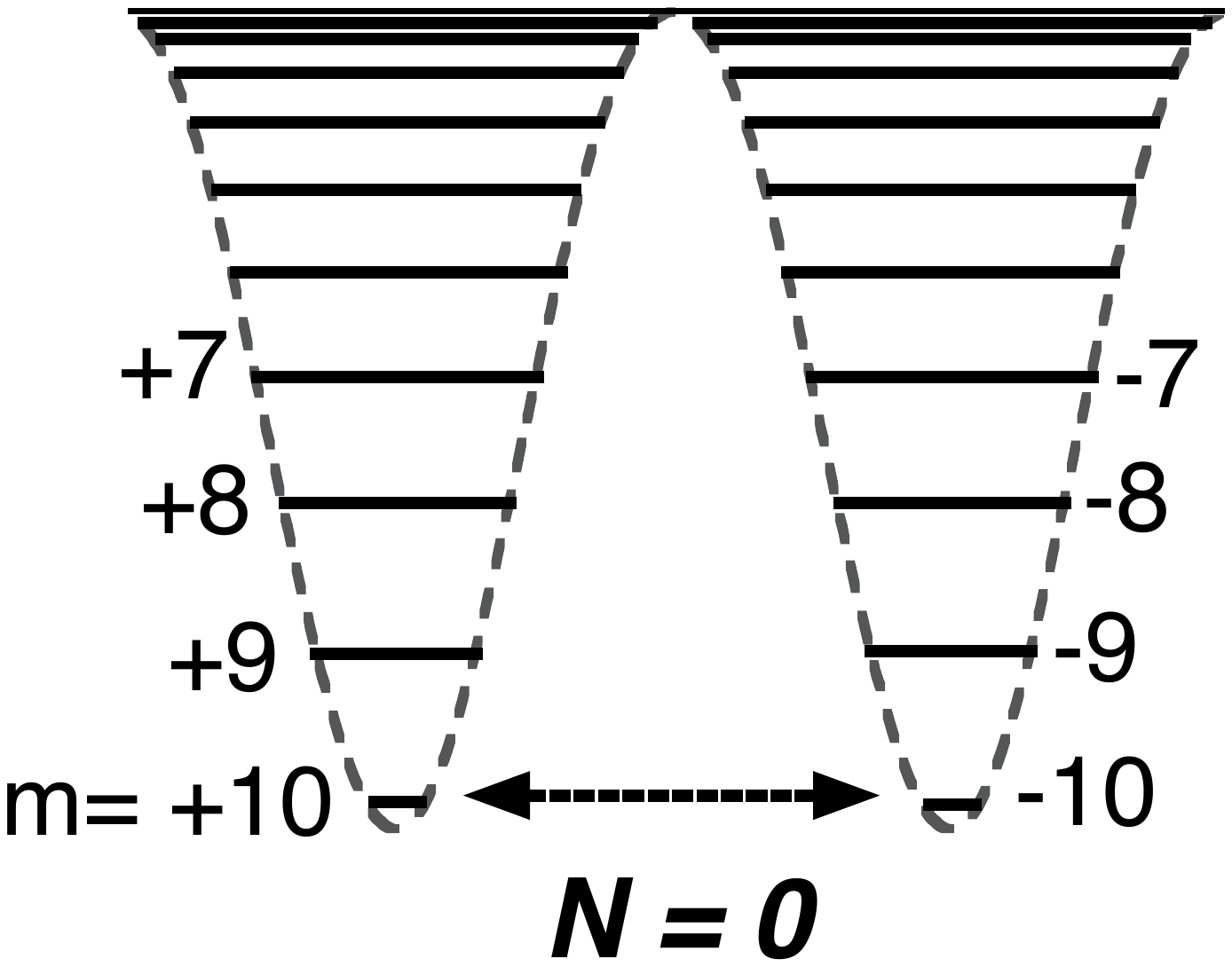}
\hspace{-0.7in}
\includegraphics[width=.39\textwidth]{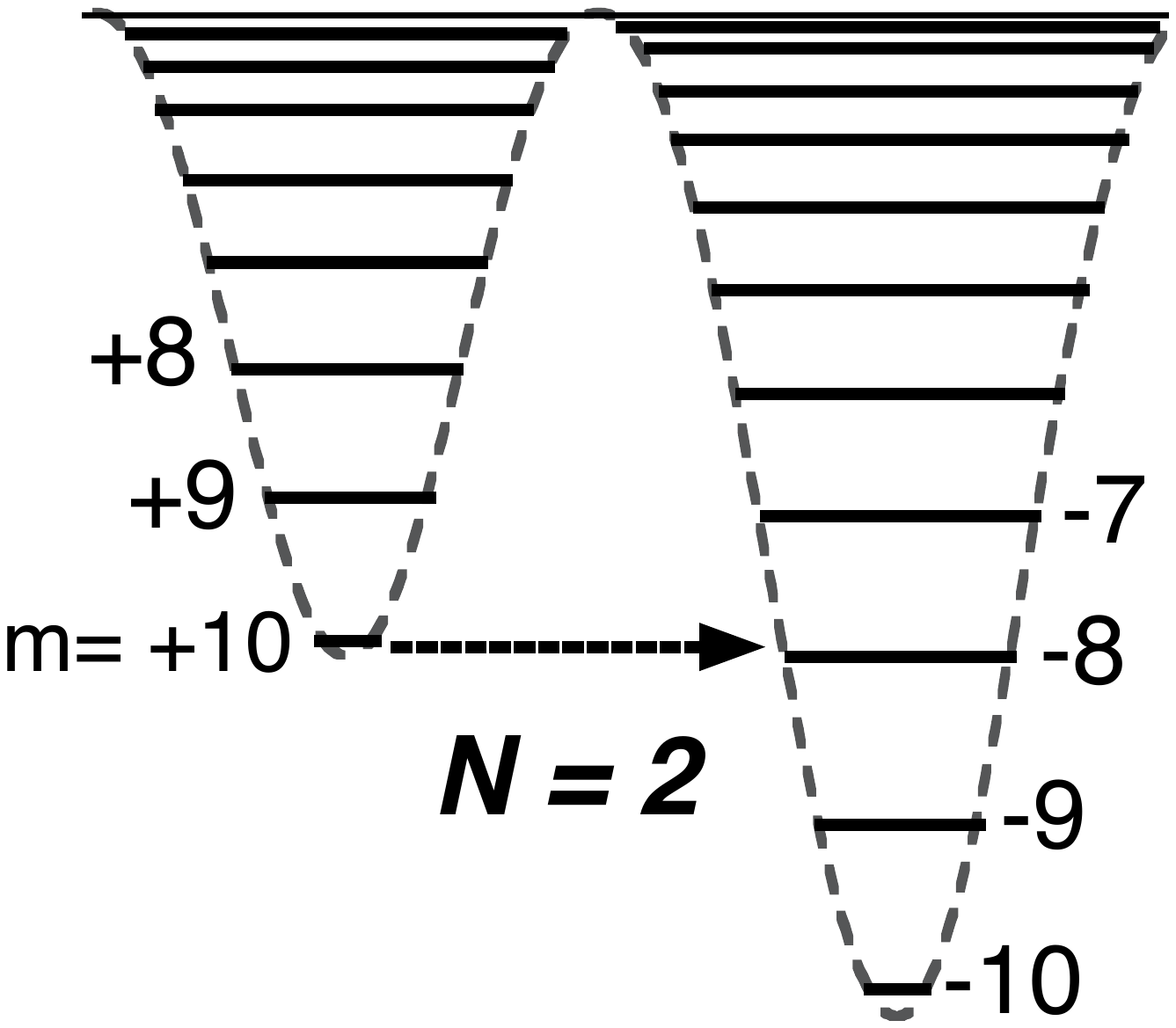}

\caption {Left panel: Chemical structure of the core of the Mn$_{12}$ molecule. The four inner spin-down Mn$^{3+}$ ions each have spin $S=3/2$; the eight outer spin-up Mn$^{4+}$ ions each have spin $S = 2$, yielding a net spin $S = 10$ for the magnetic cluster; the small grey spheres are O bridges; the arrows denote spin direction. Acetate ligands and water molecules have been removed for clarity; Middle panel: Double-well potential in the absence of magnetic field showing spin-up and spin-down levels separated by the anisotropy barrier. Different spin projection states $|m>$ are indicated. The arrows denote quantum tunneling. Right panel: Double-well potential for the N=2 step in a magnetic field applied along the easy axis.}
\label{background}
\end{figure}

As shown in the left panel of Fig. \ref{background}, the magnetic core of Mn$_{12}$-ac has four Mn$^{4+}$ (S = 3/2) ions in a central tetrahedron surrounded by eight Mn$^{3+}$ (S = 2) ions. The ions are coupled by superexchange through oxygen bridges with the net result that the four inner and eight outer ions point in opposite directions, yielding a total spin $S=10$ \cite{Sessoli}. The magnetic core is surrounded by acetate ligands, which serve to isolate each core from its neighbors in a body-centered tetragonal lattice.  A crystalline sample typically contains $\sim 10^{17}$ or more (nearly) identical, weakly interacting single molecule nanomagnets in (nearly) identical crystalline environments.

The interesting physics and potential applications of Mn$_{12}$-ac and similar materials derive from the fact that: (i) the exchange between ions within the magnetic core is very strong, resulting in a sizable, rigid spin-$10$ magnetization per molecule with no internal spin degrees of freedom at low temperatures, and (ii) the anisotropy is exceptionally large, so that the spins are bistable at low temperature, exhibiting slow relaxation and hysteresis below a blocking temperature T$_B$.  To lowest order, the spin Hamiltonian is given by:
\begin{equation}
{\cal H} = - DS_z^2 - g_z\mu_B H_z S_z + \ldots + {\cal H^\prime} 
\label{Hamiltonian},
\end{equation}
where the first term denotes the anisotropy barrier, the second is the Zeeman energy that splits the spin-up and spin-down states in a magnetic field, and the last term, $\cal H^\prime$, contains all symmetry-breaking operators that do not commute with $S_z$, thereby allowing quantum tunneling.  For Mn$_{12}$-ac, $D=0.548$K, $g_z = 1.94$; $\mu_B$ is the Bohr magneton.

As illustrated in the middle and right-hand panel of Fig.~\ref{background}, the energy is modeled as a double-well potential, with one well corresponding to the spin pointing along the easy axis in one direction and the other to the spin pointing in the opposite direction.  In zero field, there is a set of discrete, doubly degenerate energy levels corresponding to $(2S+1$) projections, $m = +10, +9,\ldots, 0, \ldots, -9, -10$, of the total spin along the easy ($c$-axis) of the crystal.  Applying a magnetic field along the easy axis lowers the energy of the potential well with spins pointing in the direction of the field relative to the potential well for spins opposite to the field.

\begin{figure}[tb]
\hspace{-.4in}
\includegraphics[width=.52\linewidth]{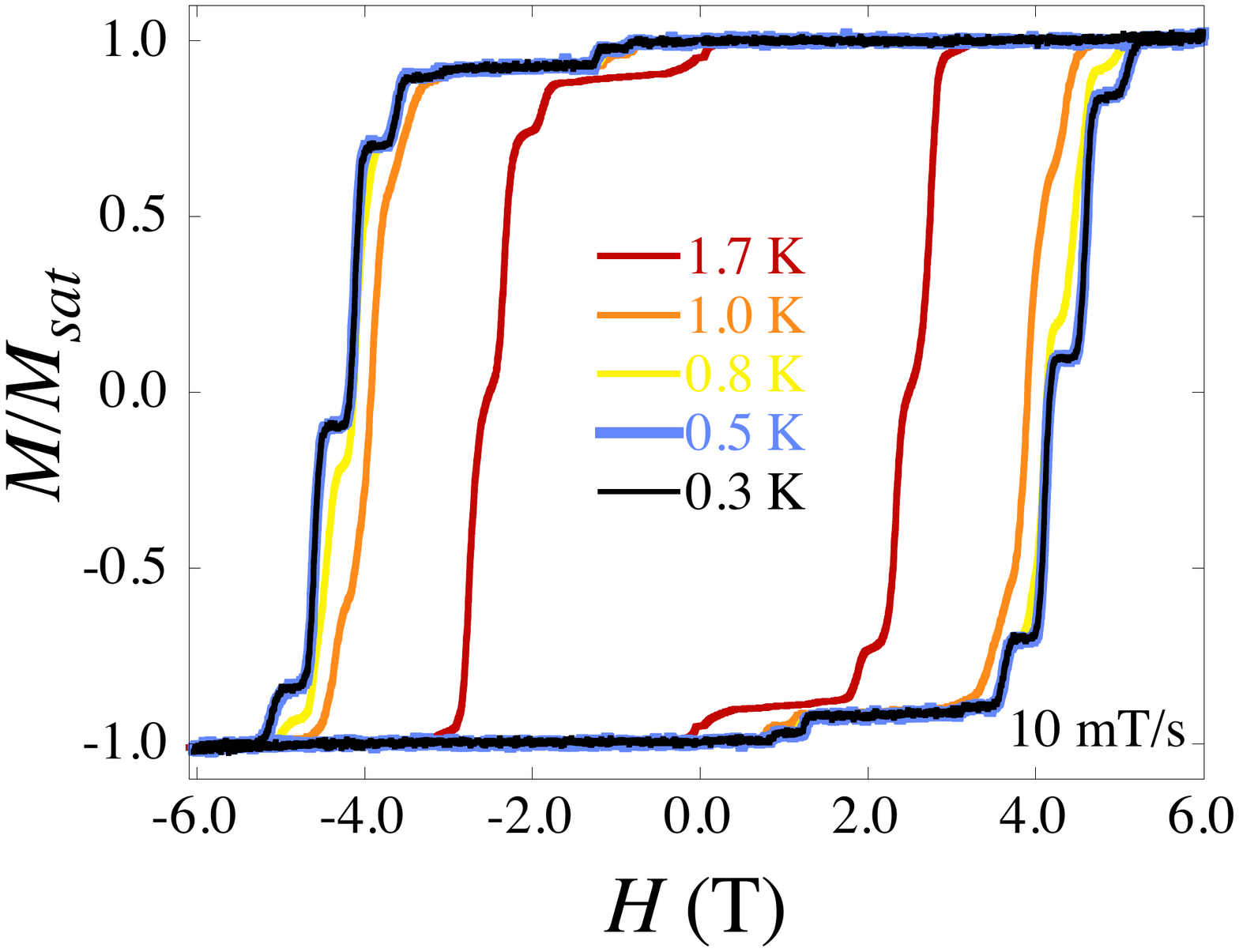}
\includegraphics[width=.5\linewidth]{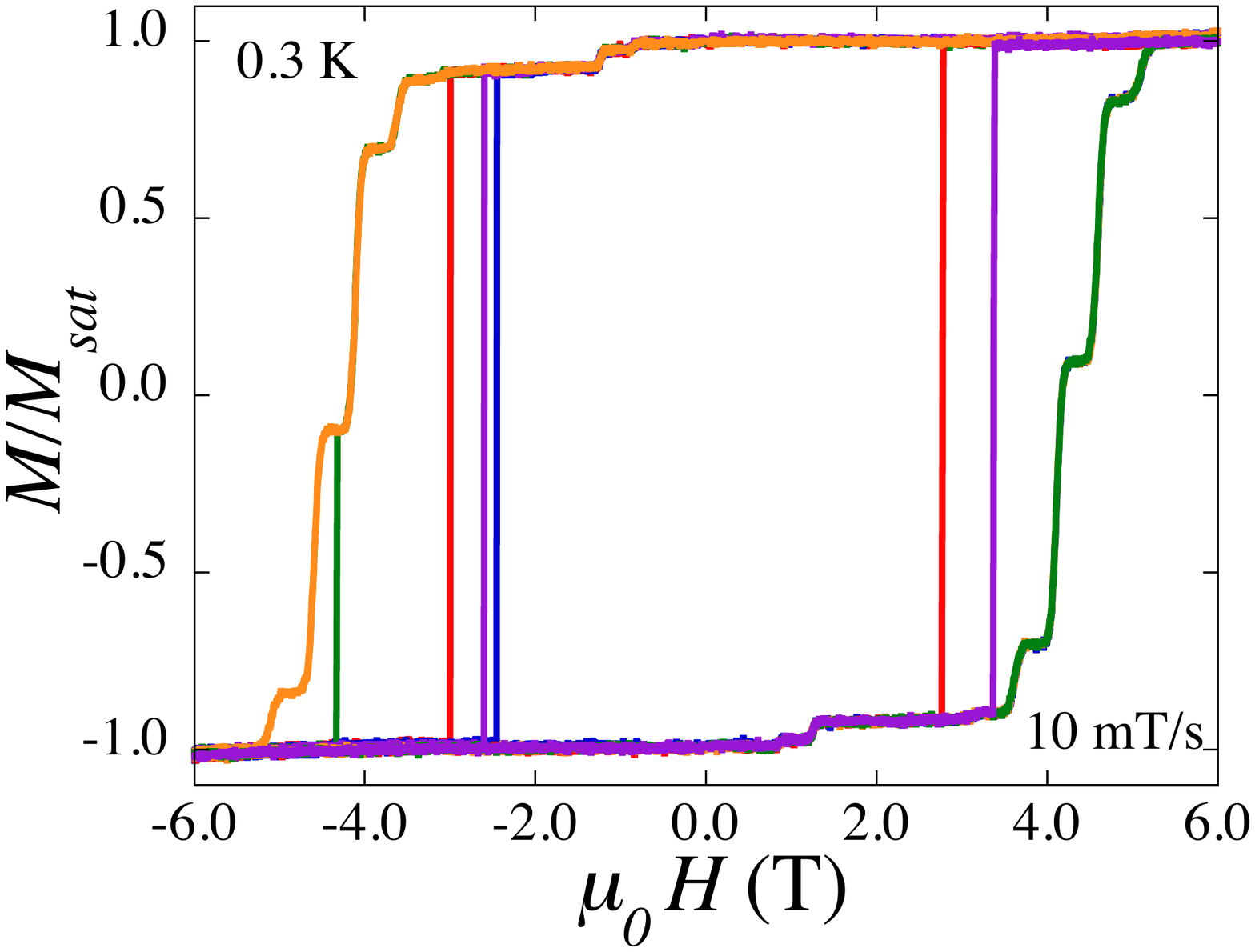}
\vspace{-,6in}
\caption{Left: Hysteresis loops of a Mn$_{12}$-ac crystal for magnetic field applied along the uniaxial $c$-axis direction at different temperatures below the blocking temperature; the magnetization is normalized by its saturation value; magnetic field was swept at 10 mT/s. Right: Hysteresis loops at $0.25$ K interrupted by magnetic avalanches (vertical lines).}
\label{loops}
\end{figure}

The relaxation rate decreases as the temperature is reduced, and below a (sweep rate-dependent) blocking temperature ($T_B \sim 3$ K), the large anisotropy barrier gives rise to slow relaxation and hysteresis loops that display steps\cite{Friedman} as a function of magnetic field $H_z$ as the magnetic field is swept from full magnetization in one direction to full magnetization in the other.\cite{Friedman,TejadaEPL,JMHernandez}  The left panel of Fig.~\ref{loops} shows the magnetization $M$ as a function of magnetic field $\mu_0H_z$.  These steps, characteristic of molecular magnets, can be understood with reference to the double well potential of Fig.~\ref{background}: faster relaxation occurs by spin-tunneling at the ``resonant'' values of the magnetic field that correspond to alignment of levels on opposite sides of the anisotropy barrier. Full saturation of the magnetization is thereby reached in a stepwise fashion, where the detailed form of the steps depend on sweep-rate and temperature.  For reviews, see Refs.~\cite{barbarareview,friedmanbook,mertesSSC,sessolireview,sessolibook,barbara2,christoureview, sarachikannual} and articles in the current volume.

\begin{figure}[h]
\hspace{-0.3in}
\includegraphics[width=2.8in]{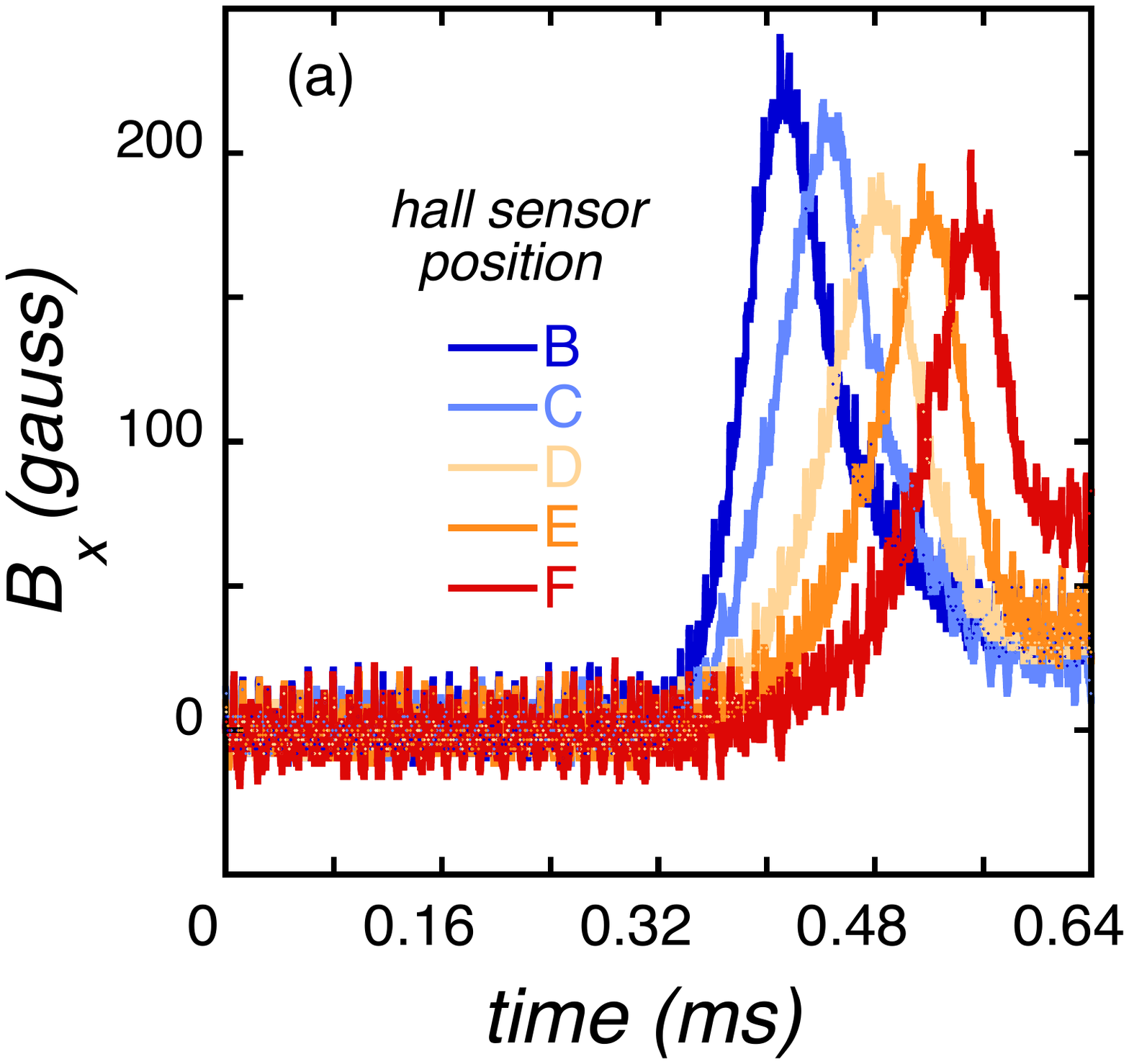}
\hspace{-0.4in}
\includegraphics[width=2.8in]{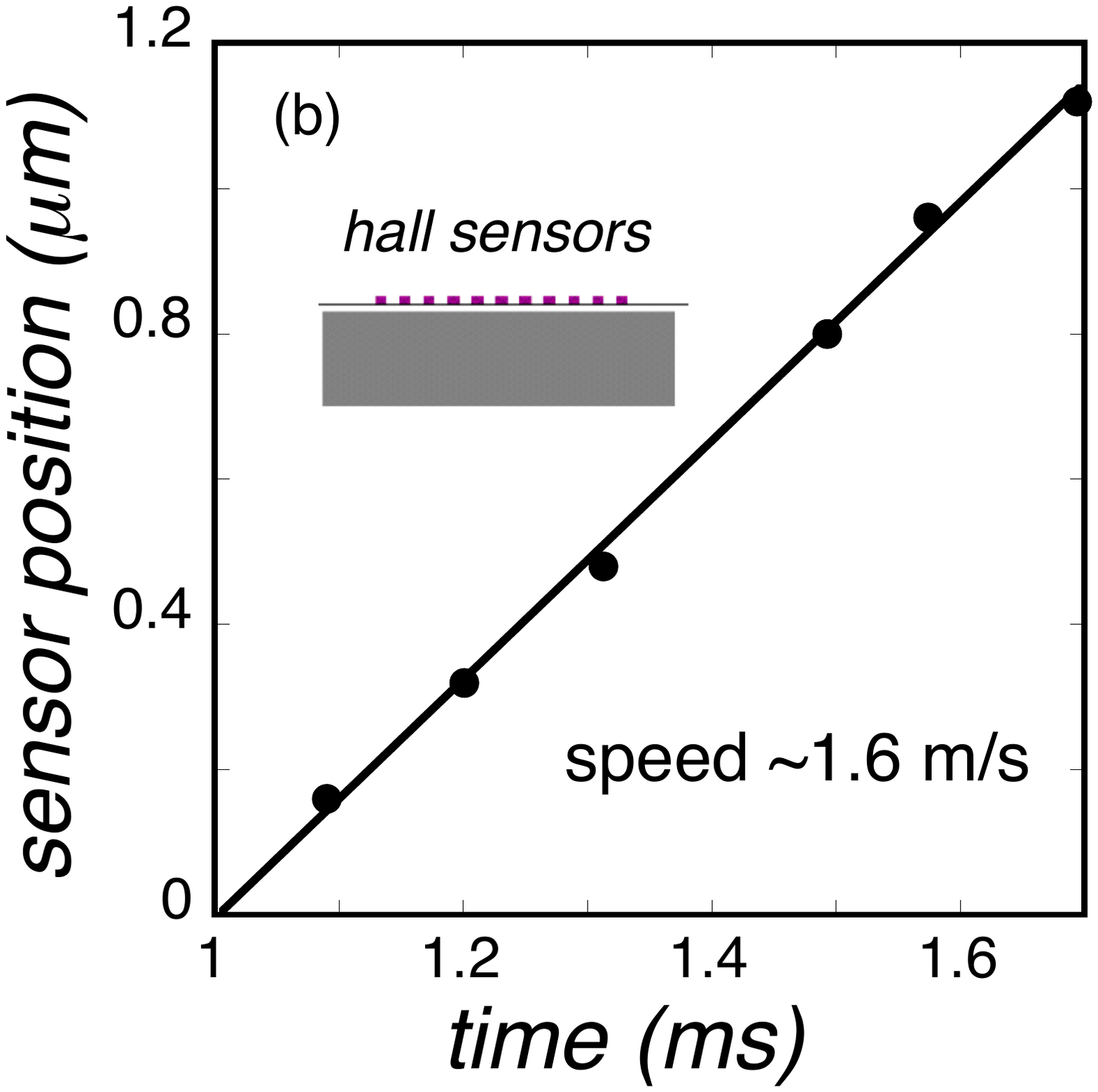}
\vspace{-.6in}
\caption{(a) The local magnetization measured as a function of time by an array of micron-sized Hall sensors placed along the surface of the sample.  The inset shows the placement of the Hall sensors on the crystal;  (b)The sensor position as a function of the time at which the sensor registered the peak.  The propagation speed for this avalanche is $2.2$ m/s, approximately three orders of magnitude below the speed of sound.  The inset illustrates the ``bunching'' of magnetic field lines as the deflagration front travels past a given Hall sensor.}
\label{deflagration}
\end{figure}

By contrast, a magnetic avalanche signals a sudden reversal of the full magnetization of the crystal, as shown in the right panel of Fig. \ref{loops}.  This process has been attributed to a thermal runaway which can be understood again with reference to the right panel of Fig, \ref{background}: when tunneling of a molecular spin occurs from the lowest state of the metastable (left-hand) well to an excited state in the stable (right-hand) well, the subsequent decay to the ground state results in the release of heat that, under appropriate conditions, can further accelerate the magnetic relaxation.  Direct measurements of the heat emitted have confirmed the thermal nature of these avalanches.

From time-resolved measurements of the local magnetization using an array of micron-sized Hall sensors placed on the surface of Mn$_{12}$-ac crystals, Suzuki {\em et al.}\cite{suzuki} discovered that a magnetic avalanche propagates through the crystal at subsonic speed in the form of a thin interface between regions of opposite spin magnetization.  Figure \ref{deflagration} (a) shows traces recorded during an avalanche by sensors placed sequentially along the easy axis near the center of a Mn$_{12}$-ac sample. Figure \ref{deflagration} (b) is a plot of the sensor position versus the time of arrival of the peak.  The inset is a schematic that illustrates the bunching of field lines at the propagating front that gives rise to the observed peaks.  From these measurements one deduces that the front separating up- and down-spins travels with a constant (field-dependent) speed on the order of $1$ to $30$ m/s, two to three orders of magnitude slower than the speed of sound.

From a thermodynamic point of view, a crystal of Mn$_{12}$ molecules placed in a magnetic field opposite to the magnetic moment is equivalent to a metastable (flammable) chemical  substance.  The release of energy by a metastable chemical substance is combustion or slow burning, technically referred to as deflagration \cite{LL}.  It occurs as a flame front of finite width propagates at a constant speed small compared to the speed of sound.  For ``magnetic deflagration'' in Mn$_{12}$-ac, the role of the chemical energy stored in a molecule is played by the difference in the Zeeman energy, $\Delta E = 2g\mu_BHS$, for states of the Mn$_{12}$-ac molecule that correspond to ${\bf S}$ parallel and antiparallel to ${\bf H}$.

The avalanches that have been studied experimentally to date are driven predominantly by the increase in temperature associated with an input of energy.  As further discussed below, Chudnovsky and Garanin have proposed a comprehensive theory to account for this process \cite{Garanin}.  In a subsequent series of papers, the same authors have pointed out that the decay rate is also affected near spin-tunneling resonances by dipolar fields that can block or unblock the tunneling \cite{cold1,cold2,cold3,cold4}.  They found that the magnetization adjusts self-consistently in such a way that the system is on resonance over a broad spatial extent, with the consequence that there can be propagating spin reversal fronts that are driven by dipolar interactions.  In general, both dipolar field and temperature are expected to control the propagation of quantum deflagration \cite{cold3,cold4}.

This review provides an overview in Section 2 of the work done to date on avalanches where temperature is the dominant driver of the deflagration front.  Section 3 briefly considers the possibility of tunneling fronts driven by dipolar interactions.  Section 4 ends the review with a brief summary and suggestions for future research.

\section{Temperature-Driven Magnetic Deflagration}

Although the probability of a spontaneous avalanche has been shown to be higher at resonant magnetic fields than off-resonance \cite{macia}, avalanche ignition is unpredictable in a swept external magnetic field, the experimental protocol that has generally been used to study the steps in the hysteresis loops.  Avalanche ignition under these conditions is a stochastic process that depends on factors such as the sweep rate, the temperature, the quality of the crystal, and perhaps other factors.  In order to carry out systematic studies of avalanche characteristics one needs to trigger avalanches in a controlled manner.  This has been achieved using a heater\cite{seandips}, and by using surface acoustic waves (which serve to heat the sample)\cite{quantumdeflagration}.  Recent studies \cite{subedi} have used current pulses.  Control of the location as well as the time of ignition could be accomplished using optical methods.

\subsection{Avalanche Ignition}

\begin{figure}[h]
\hspace{-.3in}
\includegraphics[width=.63\textwidth]{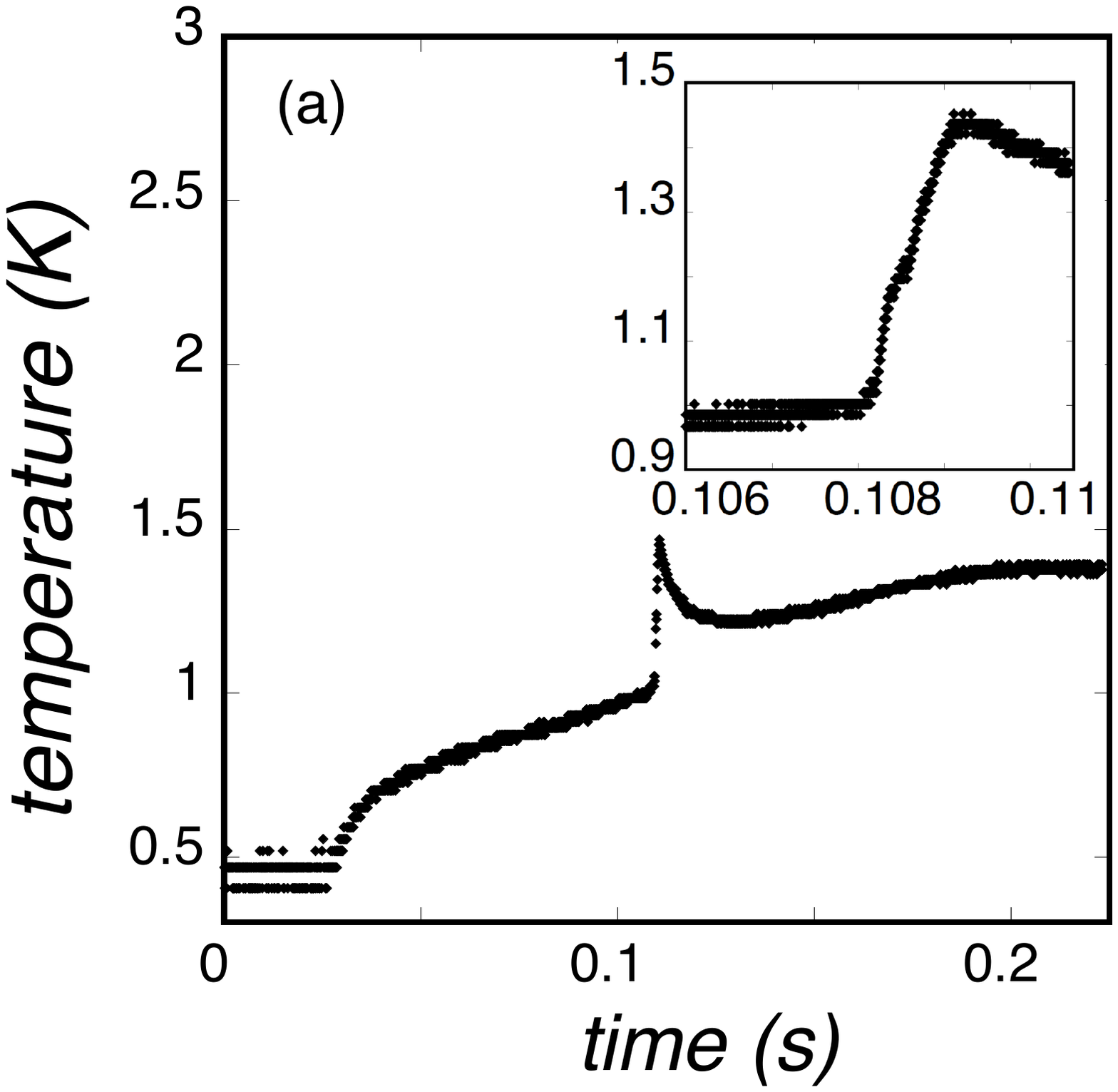}
\hspace{-.5in}
\includegraphics[width=.63\textwidth]{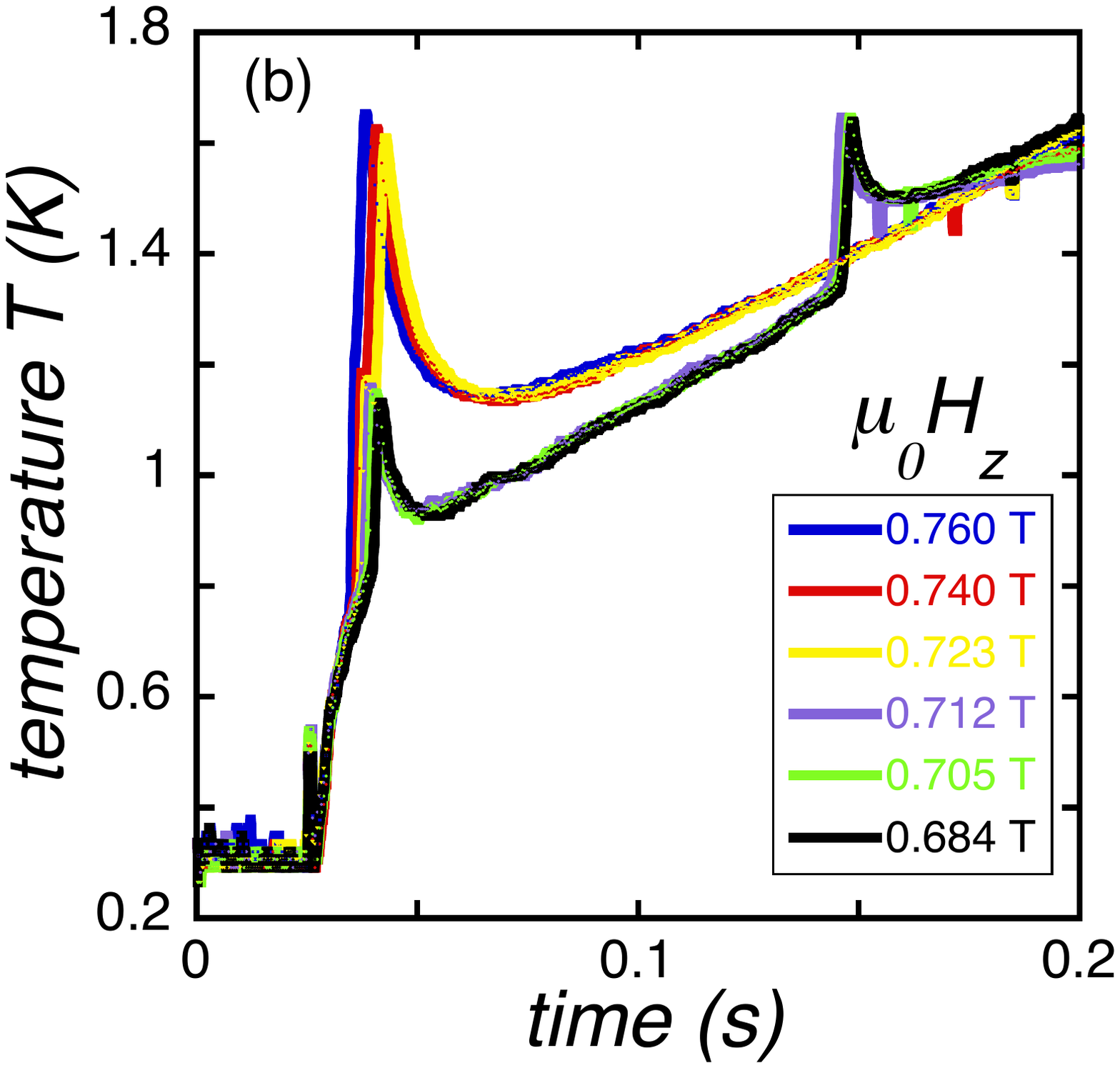}
\caption{(a) Temperature recorded by a thermometer in contact with a Mn$_{12}$ crystal during the triggering of an avalanche at $0.83$ T. The heater is turned on at $\sim 0.03$ s, the temperature then increases slowly until an abrupt rise in temperature at $0.11$ s signals the ignition of an avalanche.  The inset shows data taken near ignition with higher resolution.  The noise at low temperatures derives from digitizing  the analog output of the thermometer, which depends weakly on temperature below $0.4$ K; (b)Temperature profiles for avalanches of major and minor species triggered at low fields in a Mn$_{12}$ crystal. The two types of avalanches are triggered separately below a sample-dependent magnetic field, while at higher fields ignition of the minor species triggers the ignition of the major species.}
\label{ignition}
\end{figure}

McHugh {\em et al.}\cite{seandips} used a resistive wire element as a simple electric heater to trigger avalanches in a manner similar to the work of Paulsen and Park \cite{Paulsen}.  In these experiments, an external magnetic field is ramped to, and held at a fixed value.  The heater is then turned on to slowly heat the sample until an avalanche is triggered at a temperature measured by a small thermometer.  Avalanches launched by this method occur at well-defined, reproducible ignition temperatures.  Figure \ref{ignition}(a) shows a typical temperature profile: starting at the base temperature of $300$ mK, the temperature gradually rises  until an abrupt sharp increase in the temperature signals the ignition of an avalanche.  For this particular avalanche triggered at $\mu_0 H_z = 0.83$ T, the ignition temperature is about $1$ K.  

Single crystals of Mn$_{12}$--ac are known to contain two types of molecules. In addition to the primary or ``major'' species described at the beginning of this review,  as-grown crystals contain a second ``minor'' species at a level of $\approx 5$ percent with lower (rhombohedral) symmetry \cite{minorRef,WernsdorferI}.  These faster-relaxing molecules can be modeled by the same effective spin Hamiltonian, Eq. \ref{Hamiltonian}, with a lower anisotropy barrier of $0.49$ K.   Avalanches of each species can be studied in the absence of the other through an appropriate magnetic protocol described in Ref. ~\cite{minors}.   

Interestingly, avalanches are separately triggered by the two species in low magnetic field.  As shown in Fig. \ref{ignition}(b), at low fields the minor species relaxes prior to and independently of the major species, while above $\sim 0.7$ T the major and minor species ignite together and propagate as a single front.  It is analogous to grass and trees that can sustain separate burn fronts that abruptly merge into a single front when the grass becomes sufficiently hot to ignite the trees.
\begin{figure}[tb]
\centering
\includegraphics[width=4in]{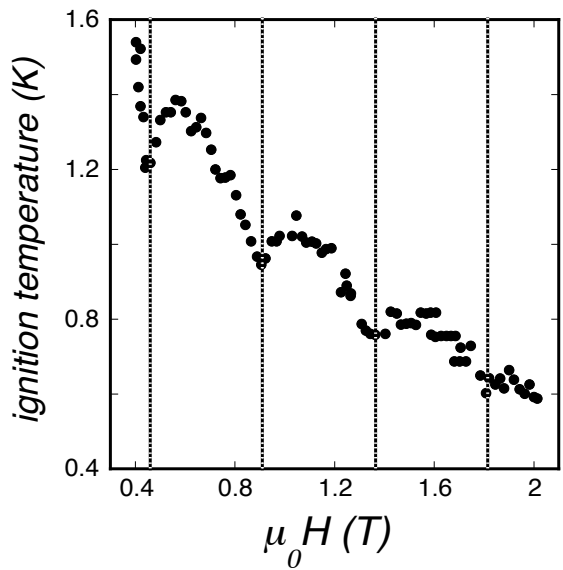}
\vspace{-1in}
\caption{Temperature required to ignite avalanches plotted as a function of magnetic field. The vertical lines denote the magnetic fields where sharp minima occur in the ignition temperature corresponding to tunneling near the top of the anisotropy barrier. The overall decrease in ignition temperature is due to the reduction of the anisotropy barrier as the field is increased.}
\label{ignitionminima}
\end{figure}

Despite the turbulent conditions that one might expect for deflagration (as in chemical combustion), quantum mechanical tunneling clearly plays a role, as demonstrated in Fig. \ref{ignitionminima}, where the temperature of ignition is plotted as a function of a preset, constant magnetic field\cite{seandips}.  The temperature required to ignite avalanches exhibits an overall decrease with applied magnetic field, reflecting the fact that larger fields reduce the barrier (see the double-well potential in Fig. \ref{background}). The role of quantum mechanics is clearly evidenced by the minima observed in the ignition temperature at the  resonant magnetic fields due to tunneling when levels cross on opposite sides of the anisotropy barrier.

\begin{figure}[h]
\centering
\includegraphics[width=.7\textwidth]{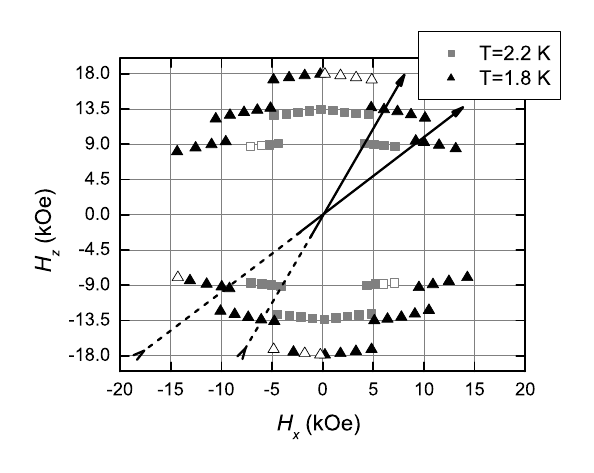}
\includegraphics[width=.7\textwidth]{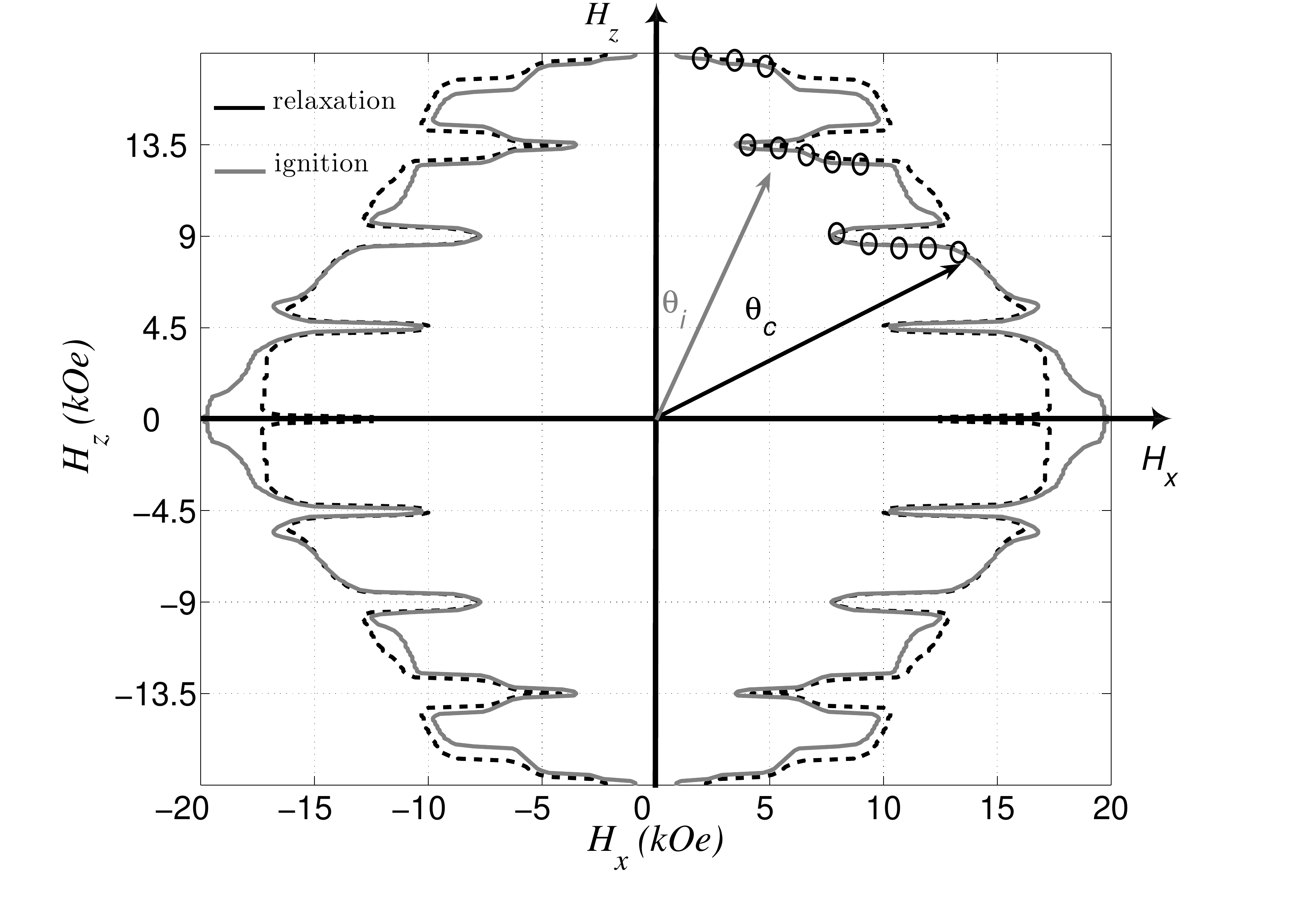}
\caption{Top: Angle dependence of metastability measured through the occurrence of avalanches.  Squares (triangles) denote parameter values where deflagration occurs for initial temperature 2.2 K (1.8 K); Bottom: Theoretical calculation for the area of stability against ignition of avalanches (solid curve) and against slow relaxation (dashed curve) \cite{note}.  Circles denote points where avalanches are predicted to occur at a given angle $\theta_i$ within the first quadrant. The angle $\theta_c$ denotes the crossing point between areas of slow relaxation and avalanche stability.  These results were obtained with T$_f$ as a parameter varying from $6.8$ K for $H = 4600$ Oe to $10.9$ K for $H = 9200$ Oe.  From Maci\`a {\em et al.} \cite{macia}.}
\label{macia}
\end{figure}

In the ignition studies described above, the barrier against spin reversal was lowered by applying a longitudinal magnetic field, $H_z$, along the uniaxial $c$-direction, which serves to unbalance the potential wells and lower the barrier against spin reversal.  Tunneling can also be promoted by applying a transverse field $H_x$ which reduces the anisotropy barrier by introducing a symmetry-breaking term, $g\mu_B H_x S_x$, to the Hamiltonian, Eq. \ref{Hamiltonian}.  Maci\`a {\em et al.}\cite{macia} investigated the threshold for  avalanche ignition in Mn$_{12}$-ac as a function of the magnitude and direction of a magnetic field applied at various angles with respect to the anisotropy axis and as a function of temperature.  As the external field is increased at a constant rate from negative saturation to positive values, both $H_z$ and $H_x$ increase, tracing a trajectory in the $(H_z,H_x)$ parameter space.  Examples of sweeps starting from zero are shown by the arrows in Fig. \ref{macia}.  An avalanche was recorded for each pair $(H_x, H_z)$ denoted by a square (for $T=2.2$ K) or a triangle (for $T=1.8$) K.

A theory of magnetic deflagration developed by Garanin and Chudnovsky \cite{Garanin} that considers only thermal effects (no dipolar interactions) explains the main features of the ignition experiments of McHugh {\em et al.} in which the critical relaxation rate was reached by varying $T_0$ with a heater, and the experiments of Maci\`a {\em et al.}, where the ignition threshold was reached by controlling the barrier $U$ using $H_x$ and $H_z$.

A deflagration front is expected to develop when the rate at which energy is released by the relaxing metastable spins exceeds the rate of energy loss through the boundaries of the crystal.  This condition can be expressed in terms of a critical relaxation rate,\cite{Garanin} 
\begin{eqnarray}
\Gamma_c = \frac{8k(T_0)k_B T_0^2}{U\langle E\rangle l^2},
\label{criticalRate}
\end{eqnarray}
where $\Gamma_c = \Gamma_0\mbox{ exp}[-U/k_BT_0]$, $T_0$ is the initial temperature, k is the thermal conductivity, $l^2$ is the characteristic cross section of the crystal, and $\langle E\rangle$ is the average amount of heat released per molecule when its spin relaxes to the stable state. The energy released when a single molecule relaxes is the Zeeman energy $\Delta E = 2g\mu_BSB_z$.  To obtain the average energy per molecule, one has to consider the fraction of molecules that relax:
\begin{eqnarray}
\langle E\rangle = 2g\mu_BS \left( \frac{\Delta M}{2M_s}\right)B_z,
\label{DeltaE}
\end{eqnarray}
where $M_s$ is the saturation magnetization and $\Delta M = |M_z - M_s|$ is the change from initial to final magnetization.

Calculations based on Eq. \ref{criticalRate} yield the curves shown in Fig. \ref{macia} (b). Two areas are defined in the $(H_z,H_x)$ parameter space where the spins are expected to be metastable against relaxation: the solid line denotes the region of metastability against relaxation by triggering avalanches while the dashed curve delineates the region of metastability against slow, stepwise relaxation \cite{note}.  If the experimental trajectory, denoted by the arrows, crosses the grey solid line first, an avalanche will ignite.  If the dashed line is crossed first, the metastable spins will relax slowly without triggering an avalanche.  This defines a critical angle $\theta_c$, above which an avalanche cannot occur.

Maci\`a {\em et al.} measured the ignition threshold by applying an increasing external field at an angle with respect to the crystal.  The relaxation rate increases as the field grows until $\Gamma_c$ is reached and deflagration ignites, as shown in Fig.~\ref{macia} (a).  For sufficiently large values of $H_x$, they found that the slow relaxation of the metastable spins occurs before deflagration can ignite.  This defines a line in parameter space separating regions where one or the other mode of relaxation occurs, as shown in Fig. \ref{macia}(b).  The theory predicts that the transverse field should result in a significant decrease in the magnetization metastability at the resonant fields of $H_z$.  The data recorded in Fig. \ref{macia}(a) confirm this and are consistent with the ignition temperatures of Fig. \ref{ignitionminima}.  In addition, ignition thresholds were measured at two different temperatures.  The area of stability is clearly reduced by the increased initial temperature, as expected.

\subsection{Avalanche Speed}

\begin{figure}[tb]
\centering
\includegraphics[width=4in]{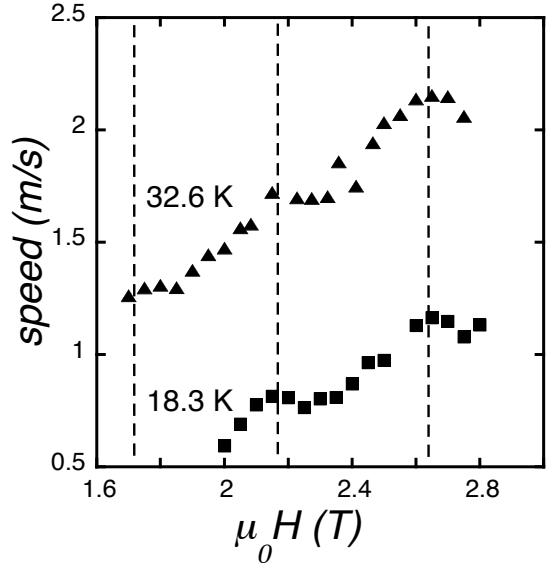}
\vspace{-1in}
\caption{The speed of propagation of the magnetic avalanche deflagration front as a function of the (fixed) field at which the avalanche is triggered.  Data are shown for category $C$ avalanches for which the average energy released, $\langle E\rangle$, is held constant at $18.3$ K and $32.6$ K (see text).  Note the enhancement of propagation velocity at magnetic fields corresponding to quantum tunneling (denoted by vertical dotted lines).  From McHugh thesis \cite{mchughthesis}}
\label{speedmaxima}
\end{figure}

Hern\'andez-M\'inguez {\em et al.}\cite{quantumdeflagration,TejadaJMMM} carried out a systematic investigation of avalanche speeds as a function of a preset, constant magnetic field $\mu_0 H_z$ for avalanches triggered by surface acoustic waves.  From SQUID-based measurements of the total magnetization of a crystal of known dimensions, and the realization that the avalanche propagates as an interface between regions of opposite magnetization \cite{suzuki}, they deduced that the speed of propagation of the avalanches is enhanced at the resonant fields where tunneling occurs, confirming the important role of quantum mechanics and prompting the authors to name the phenomenon ``quantum magnetic deflagration."  Similar results were obtained from local, time-resolved magnetization measurements using micron-sized Hall sensors \cite{mchughthesis}, as shown in Fig. \ref{speedmaxima}.

McHugh et al.\cite{mchugh2} reported a detailed, systematic investigation of the speed of magnetic avalanches for various experimental conditions.  The speed of propagation of an avalanche is described approximately\cite{suzuki} by the expression, $v\sim  \kappa/\tau_0)^{1/2} exp{[-U(H)/2k_BT_f]}$, where $U$ is the barrier against spin reversal, $T_f$ is the flame temperature at or near the propagating front where energy is released by the reversing spins,  $\kappa$ is the thermal diffusivity, and $\tau_0$ is an attempt time.  We note that the energy barrier $U$ and the flame temperature $T_f$ appear only as the ratio $U/T_f$ in the above expression for the velocity.  It is therefore convenient to plot the speed of the avalanche as a function of $U/T_f$.

In the studies of McHugh et al.\cite{mchugh2}, avalanches were controllably triggered using three different protocols, as follows:
 
(A) From fixed (maximum) initial magnetization in various external fields; there is full (maximum) magnetization reversal, $\Delta M/2M_s = 1$; both $U$ and $T_f$ vary; 

(B) In fixed external field starting from different initial magnetization; here the amount of ``fuel'' $\Delta M/2M$ is varied for a fixed magnetic field (thus $U$ is held constant); the avalanches differ primarily through the amount of energy released - the flame temperature $T_f$ varies; 

(C) for fixed energy release, thus fixed $T_f$, by adjusting external magnetic fields and initial magnetization.\\

\begin{figure}[tb]
\vspace{-.3in}
\hspace{-0.4in}
\includegraphics[width=.63\textwidth]{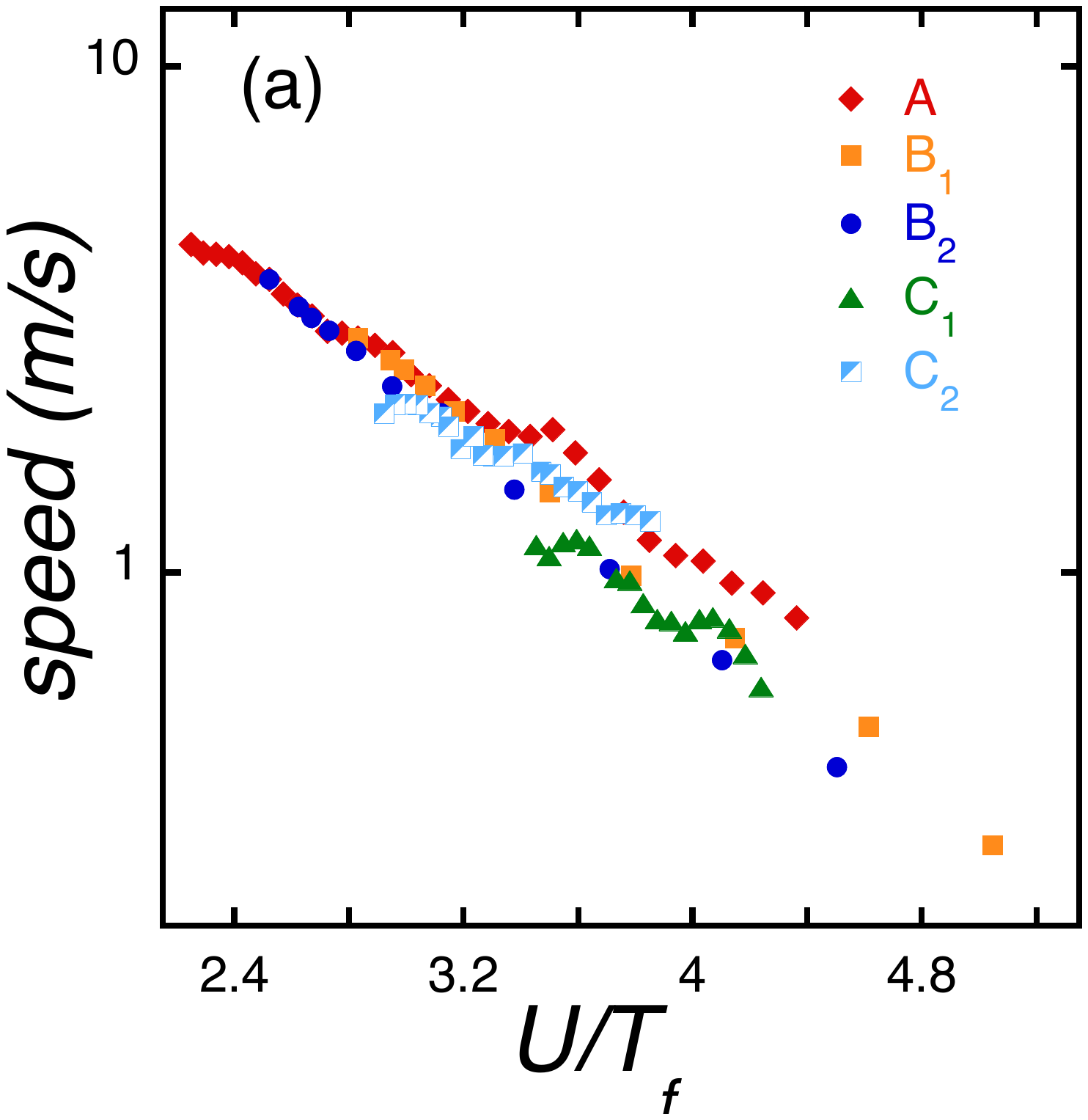}
\hspace{-0.6in}
\includegraphics[width=.63\textwidth]{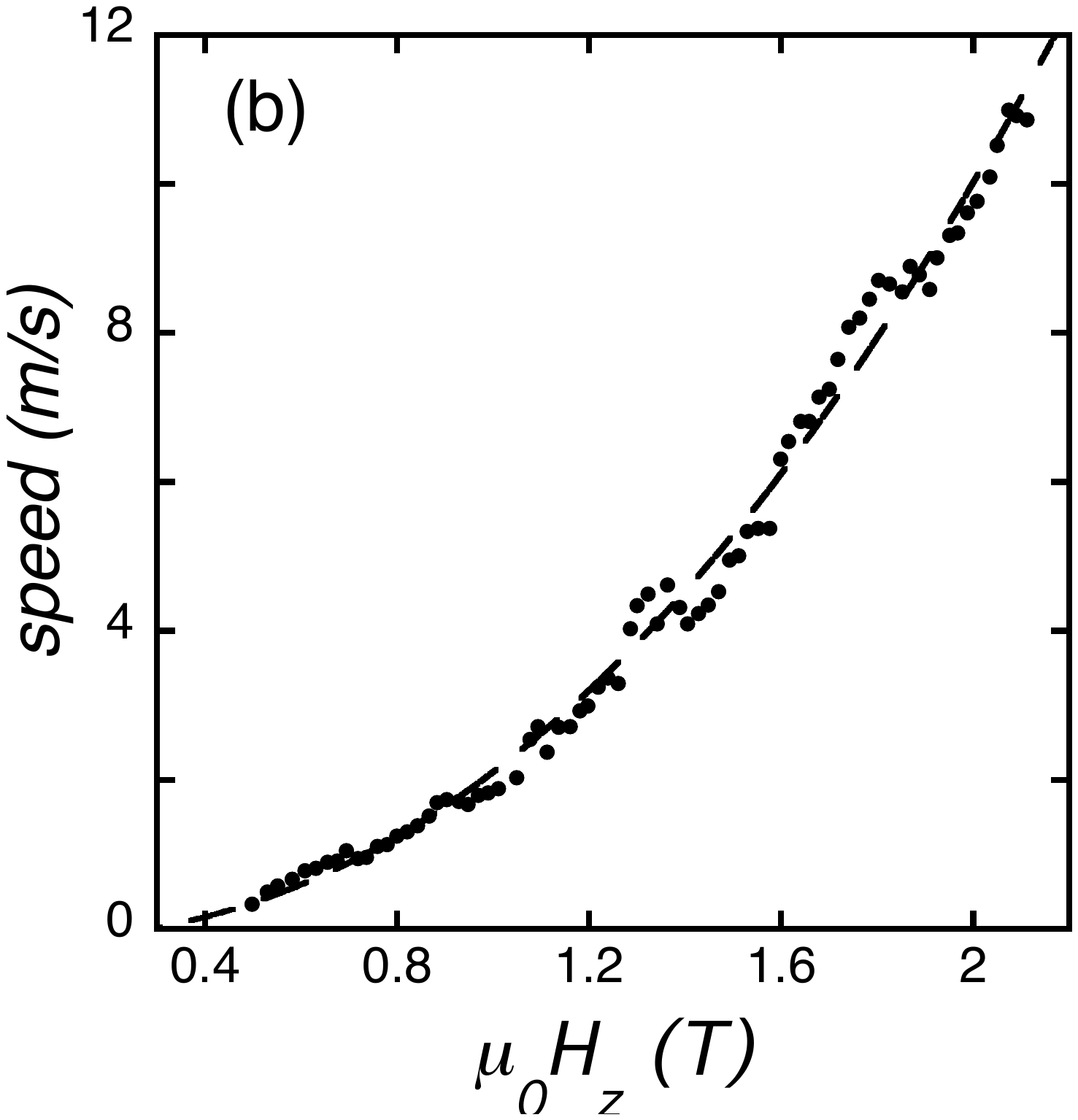}
\caption{(a) Avalanche speeds for a single crystal with various initial magnetic preparations.  $A$ denotes avalanches with $\Delta M/2M_s = 1$; $B_1$ and $B_2$ denote data taken at $\mu_0H_z = 2.2$ T  and $2.5$ T, respectively;  $C_1$ and $C_2$ denote avalanches with estimated flame temperatures $T_f \approx 10$ K  and $12$ K, respectively.  (b) Avalanche speeds for different crystal with $\Delta M/2M_s = 1$.  The fit requires an unphysical temperature dependence for the thermal diffusivity, $\kappa \propto T^{3.5}$}
\label{tuning}
\end{figure}

The theory of magnetic deflagration \cite{Garanin} provides the following theoretical expression for the speed of the deflagration front: 
\begin{eqnarray}
v = \sqrt{\frac{3k_BT_f\kappa \Gamma(B, T_f)}{U(B)}}.
\label{OneDSpeed}
\end{eqnarray}
If one assumes the thermal diffusivity $\kappa$ is approximately independent of temperature, or that its temperature dependence is unimportant compared to that of other parameters in the problem, then all measured avalanche velocities should collapse onto a single curve when plotted as a function of $(U/T_f)$.

Figure \ref{tuning} (a) \cite{mchugh2} shows the measured avalanche velocity as a function of $(U/T_f)$ obtained using the three different protocols described above.  Although an approximate collapse is obtained, there are clear and systematic deviations.  That these different experimental protocols introduce systematic variations, albeit small, suggests that the theory is incomplete.

Shown in Fig. \ref{tuning} (b), an attempt to fit to the theory by allowing the thermal diffusivity to vary as a power law of the temperature for avalanches of type (A) that involve full magnetization reversal yields $\kappa \sim T^{3.5}$.  This is a distinctly unphysical result, as the thermal diffusivity is generally a strongly decreasing function of temperature \cite{LowTemperaturePhysics} for these materials.  We note that experimental measurement of the thermal diffusivity of Mn$_{12}$ are not available.

In brief, the Chudnovsky-Garanin theory of deflagration captures the main features found in the experiments.  However, although the ignition experiments have yielded results that agree with it in detail, the theory does not provide a fully satisfactory description of the speed of propagation of the deflagration fronts.  It is possible that dipolar interactions (discussed in the next section) play a sufficiently important role to account for the discrepancies.
 
\section{Cold Deflagration}

Although small compared to other energies, dipolar interactions are now recognized as playing an important role in many molecular magnets. This is confirmed by reports of ferromagnetism mediated by dipolar interactions below 1 K in Mn12-ac \cite{luis,RFIFM}, as well as in other molecular magnets \cite{morello,evangelisti,burzuri}.  Even in the paramagnetic phase, where long range order is not realized, the change of the spin state of a molecule results in a change of the long-range dipolar field acting on other nearby spins \cite{garanindipolar, liu}.  Thus, dipolar interactions can tune spins in and out of resonance and can thereby have a profound influence on the spin dynamics.

D. A. Garanin and E. M. Chudnovksy \cite{cold1,cold2,cold3,cold4} have proposed that propagating fronts of spin reversal (avalanches) can occur that are driven by the dipolar interactions between magnetic molecules. Their numerical simulations show that dipole-dipole forces establish spatially inhomogeneous states in molecular magnets \cite{avraham} such that there is a self-consistent adjustment of the metastable population acting to create a dipolar field that is constant over a sizable region of the sample, thereby bringing the system to resonance over an extended region where all the spins can relax collectively by tunneling.  This, in turn, can lead to propagating fronts of spin reversal, which they have dubbed ``cold deflagration''.  

Interestingly, Garanin and Chudnovsky have noted that such collective traveling spin reversal fronts are potential sources of Dicke superradiance \cite{superradiance,calero,henner,yukalov,yukalov2,benedict,tokman} at frequencies in the teraHertz range, a particularly interesting region of the electromagnetic spectrum where few sources are available.  If self-organization does result in a uniform dipolar field within the tunneling front, the resonant condition is fulfilled for a macroscopic number of magnetic molecules inside the front, and it is indeed plausible that these avalanches could emit a superradiant electromagnetic signal.  Intense bursts of radiation have indeed been detected experimentally during magnetic avalanches. There has been much speculation that this could be Dicke superradiance, but experiments have been inconclusive on this very interesting issue \cite{super1,super2,super3,super4,super5,keren}.

The avalanches that have been studied experimentally to date have been triggered in large longitudinal bias fields near the higher-number field resonances.  In these circumstances, the spins tunnel from a metastable state and decay to a ground state of opposite spin that is much lower in energy, releasing Zeeman energy to the phonon system and generating heat.  This triggers thermal avalanches, as confirmed by a measured increase in the temperature of the crystal.  It will be of great interest to find magnetic avalanches that are driven (or partlially driven) by dipole-dipole interactions, and to study the relative roles of cold and ``hot'' deflagration for different parameters (temperature, parallel and perpendicular magnetic field, sweep rate, and so on).  The possibility that superradiance will be emitted in the process is particularly exciting.

Although hints of cold deflagration may have been found, there are no definitive reports of this process to date.  We note that a particularly large effect is expected in the presence of a strong transverse field which promotes tunneling and lowers the anisotropy barrier, so that relaxation toward equilibrium can proceed by tunneling at zero longitudinal bias field without thermal assistance and without releasing Zeeman energy into the system.  However,  recent experiments \cite{subedi} show that in a strong transverse magnetic field the relaxation near tunneling resonances becomes so rapid that it is difficult to create an initial state with a sizable out-of-equilibrium population sufficient to trigger a tunneling front. This is a major experimental challenge for realizing dipole-driven spin-reversal fronts.

\section{Summary and Outlook for the Future}

Once considered events to be avoided, as they interfere with a detailed study of the stepwise process of magnetization via spin-tunneling, magnetic avalanches have recently been the focus of attention and renewed interest, partly stimulated by the theoretical suggestion that the radiation emitted during an avalanche may be in the form of coherent (Dicke) superradiance\cite{superradiance}. Although the issue of coherence of the radiation has yet to be resolved, recent studies have clarified the nature of the avalanche process itself.

Magnetic avalanches proceed as traveling fronts along which the molecules reverse their spin, releasing Zeeman energy which drives the spins to reverse throughout the crystal.  These spin-reversal fronts  propagate at subsonic speeds, and are  analogous to the process of chemical combustion, technically known as chemical deflagration: here a chemical reaction propagates along a front where energy is released that drives the reaction front at subsonic speed.  A burning sheet of paper is a clear example of chemical deflagration.  A great advantage of the magnetic analog is that, unlike burning paper, it is non-destructive, fully reversible and continuously tunable using an external magnetic field.  Magnetic deflagration is thus amenable to carefully controlled study.

In this chapter, we have reviewed experiments on the ignition and the speed of propagation of a magnetic avalanche driven by the release of Zeeman energy at the deflagration front.  The conditions for ignition and the speed of propagations both show clear effects of quantum mechanics at the resonant fields that allow tunneling across the anisotropy barrier.  The theory of magnetic deflagration of Chudnovsky and Garanin is in excellent agreement with the parameters determined experimentally for ignition.  The theory also provides a good qualitative fit to the observed avalanche velocity, but there are detailed discrepancies that suggest that additional factors need to be included to obtain good quantitative agreement.  The effect of dipolar interactions must clearly be included in a full theory \cite{cold4}.

Dipole-dipole interactions are sufficiently strong in some molecular magnets that they lead to long-range ordering at low temperatures.  A particularly interesting consequence of dipolar interactions is ``cold deflagration'' proposed by Garanin and Chudnvosky, a process in which spin-reversal fronts are driven predominantly by dipolar effects.  These have not yet been realized (or perhaps recognized) experimentally.  Garanin and Chudnovsky suggest that self-organization of the internal dipolar fields brings molecules into resonance over a broad front that may serve as a source of coherent teraHertz radiation.  In addition to the intrinsic interest, it would be interesting to search for cold deflagration as a potential source of Dicke superradiance in this difficult and important range of the electromagnetic spectrum, where few sources are available. 

More experimental work is clearly needed.  Currently underway, a detailed investigation of avalanche ignition in fixed transverse and fixed longitudinal (bias) field is yielding interesting, new results \cite{subedi}.  Measurements of the thermal diffusivity would provide an important constraint on the theory, as would a reliable determination of the (local) temperature of the deflagration front.  Investigations of the influence of sample shape, size and quality would also be illuminating.  Spatial control of the avalanche ignition points, possibly by optical means, could provide important information.  Studies of the shape of the deflagration front, and its character (turbulent or laminar) would be particularly interesting.

The possibility of observing a transition to detonation is intriguing \cite{bychkov, cold4}.  Deflagration is but one type of combustion process.  Another, more violent type, is detonation, where heat spreads from the reaction front as a shock wave rather than by diffusion.  It is natural to ask whether crystals of molecular magnets can support the magnetic analog of chemical detonation.  Decelle et al.\cite{Decelle} have reported results hinting at this possibility using high external field sweep rates ($4$ kT/s).  The interpretation of these experiments is not entirely clear, and much work remains to be done.  

We close by noting once more that, to the degree that magnetic deflagration resembles chemical deflagration, the magnetic manifestation of this process offers some clear advantages for the study of chemical combustion.  The magnetic analog is non-destructive and reversible, enabling a broad range of controlled studies on a single sample.  Unlike the chemical process, it is a particularly interesting realization of deflagration in which quantum mechanical tunneling plays an important role.

\section*{Acknowledgements}

The author thanks the students whose experiments made this review possible, most particularly Yoko Suzuki and Sean McHugh; and Eugene Chudnovsky and Dmitry Garanin for their careful reading of this manuscript.  Support was provided by NSF Grant No. DMR-00451605.


\begin{thebibliography}{5}
\bibitem{note} As is true for the determination of blocking temperatures, the position of the dashed line depends on the typical experimental time scale.
\end{thebibliography}
\end{document}